\documentclass[twocolumn,aps,showpacs]{revtex4}

\usepackage{amsbsy}
\usepackage{amsmath}
\usepackage{epsfig}
\usepackage{amsfonts}
\usepackage{graphicx}
\usepackage{latexsym}

\newcommand{\be}{\begin{equation}}
\newcommand{\ee}{\end{equation}}
\newcommand{\bea}{\begin{eqnarray}}
\newcommand{\eea}{\end{eqnarray}}

\begin{document}

\author{Li-Xiang Cen$^1$ and Paolo Zanardi$^{1,2}$}
\affiliation{$^1$Institute for Scientific Interchange Foundation, Viale Settimio
Severo 65, I-10133 Torino, Italy\\
$^2$Department of Mechanical Engineering, Massachusetts Institute of Technology, 
Cambridge, Massachusetts 02139}
\title{Decoherence suppression for oscillator-assisted geometric quantum gates via 
symmetrization}

\begin{abstract}
We propose a novel symmetrization procedure to beat decoherence for 
oscillator-assisted quantum gate operations. The enacted symmetry is related to 
the global geometric features  of qubits transformation
based on ancillary oscillator modes, e.g. phonons 
in an ion-trap system. It is shown that the devised multi-circuit 
symmetrized evolution endows the system with a two-fold resilience against 
decoherence: insensitivity to thermal fluctuations and quantum dissipation.
\end{abstract}


\maketitle

A key requirement to implement quantum computation consists in the ability 
to perform, with high accuracy, some set of universal gate operations for 
quantum bits (qubits) 
\cite{QC}. In the real world, a quantum system is never isolated, and the 
decoherence due to the surrounding
environment is regarded as the main obstacle to practical realization of 
quantum information processing.  A variety of techniques,
based on the underlying mechanism of symmetry, have been proposed to
protect quantum information, including quantum error correcting codes  
\cite{shor}, decoherence-free subspaces \cite{EAC}, bang-bang dynamical 
decoupling \cite{bang}, and noiseless subsystems \cite{knill00,stab}.

Most of these strategies rely, in a way or the other, on symmetry 
properties
of the dynamics of the plain interactions of quantum information manipulation 
\cite{stab}. For instance, in the proposed schemes for dynamical decoupling 
\cite{bang}, the qubit-qubit coupling is often invariant with respect to the 
symmetrization procedure exploited for the elimination of the qubit-environment 
couplings. On the other hand, it has been recently argued that robust quantum computing 
might be also achieved by a class of geometric transformations \cite{zanardi}.
In particular, geometric quantum manipulation seems to possess a certain intriguing 
resilience to classical noises, i.e. stochastic fluctuations of the control parameters, 
due to its inherent global feature of the operation \cite{solinas-zhu}.

In this paper, we shall contrive a novel symmetrization procedure to beat
decoherence for a sort of geometric quantum gates based on the use of
ancillary oscillator modes. For the typical ion-trap system, the global gate 
operation assisted by a phonon mode \cite{milburn} and the related time-reversal 
symmetry will be exploited. At variance with  the previous schemes, the symmetry
involved in our proposal will not directly show up at the level of dynamical interactions,
i.e. Hamiltonian operator, but will relate to the geometric nature of the overall, 
finite-time evolution, i.e. unitary operator.
The use of this sort of global transformations to implement quantum
gates, and their extra advantages, have been already discussed by several 
authors \cite{molmer}. However, the symmetry associated with these geometric operators
and its potential relevance to the struggle against decoherence was never revealed.

Let us focus, for the sake of concreteness, on quantum information processing based on
ion-trap systems. Quantum manipulation on this kind systems constitutes one of 
the most promising approaches to quantum computation and it has been intensively studied
during the past years \cite{molmer,cirac,it-others}. In detail, qubits in the system are
stored in internal atomic states controlled by laser beams and a controlled interaction
between them is obtained via the coupling with ion vibrational states.
Although the coupling of the qubits with the phononic degrees of freedom will
generally lead to dependence of the gate operation on the motional states,
recent studies show that such a drawback can be avoided by resorting
to a control process enacting a gate with global features \cite{milburn,molmer}.
The main idea of the scheme can be illustrated as below. Suppose that, by
appropriately choosing the laser intensities and detunings, the following
coupling of two ion qubits with the oscillator degree of freedom can be obtained 
\begin{equation}
H(t)=[\alpha (t)a+\alpha ^{*}(t)a^{\dagger }]J_y,  \label{hamil}
\end{equation}
where the collective Pauli operator $J_y=\sigma _y^{(1)}+\sigma _y^{(2)}$
accounts for the internal states of the ion qubits and $\alpha (t)$
characterizes the interaction of the qubits with the vibrational
state. When  $\alpha (t)$ has the form of certain periodic function such that
in the time interval $\int_0^T\alpha (t)dt=0$, the corresponding evolution
generates a global transformation, $e^{-iA(T)J_y^2}$ with $%
A(T)=i\int_0^T\alpha (t)\int_0^t\alpha ^{*}(t^{\prime })dt^{\prime }dt$,
which acts {\em only} on the internal states of the ions. The simplest case is
given by the periodic step pulses, which leads to the following gate
transformation 
\begin{equation}
U_C(T)=P_{+}^x(\tau )P_{+}^p(\tau )P_{-}^x(\tau )P_{-}^p(\tau )=e^{-i2\alpha
_0^2\tau ^2J_y^2},  \label{pulse}
\end{equation}
where $\alpha _0$ and $\tau =\frac 14T$ stand for the pulse amplitude and length
respectively; $P_{\pm }^x(\tau )=e^{\pm i\sqrt{2}\alpha _0\tau xJ_y}$ and 
$P_{\pm }^p(\tau )=e^{\pm i\sqrt{2}\alpha _0\tau pJ_y}$ describe
the unitary transformations for corresponding pulse interactions. [We have
used the notation $x=\frac 1{\sqrt{2}}(a+a^{\dagger })$ and $p=\frac i{\sqrt{
2}}(a^{\dagger }-a)$]. Note that the transformation (\ref{pulse}) contains no
operators with a non-trivial  action over the vibrational degrees of freedom.
The advantage of this  gate operation is evident:  it is independent of the vibrational 
states, hence is insensitive to the ion temperature.

One more feature of the above described  operation, which will be
at the very basis of  the scheme to beat decoherence proposed in this paper, 
is the invariance of the global action under the {\em time reversal} of the interaction:
$ H_C(t)\rightarrow H_{\bar{C}}(t)=-H_C(t)$. This property can be easily checked
in the case of the step pulse interaction. Indeed one has:  $U_{\bar{C}}(T)=
P_{-}^x(\tau )P_{-}^p(\tau )P_{+}^x(\tau )P_{+}^p(\tau )=U_C(T)$.
Represented in Fig. 1 are the corresponding circuits in the oscillator phase space, 
including the above step pulse interaction and also the circular case with $\alpha
(t)=\alpha _0\cos \frac{\pi t}{2\tau }+i\alpha _0\sin \frac{\pi t}{2\tau }$.
The invariance of the global action under time reversal has, of course,
a simple geometric interpretation: the corresponding paths traversed in the
phase space, being related by the map $\alpha\rightarrow -\alpha,$ indicate
an anti-symmetric track geometry.

In real-life systems, the qubits and the oscillator degrees of freedom
cannot be perfectly isolated. Here we shall focus on the case in which 
decoherence effects will arise by a thermal dissipation of the 
oscillator mode only. 
This latter will spoil the global action of (\ref{pulse}) and induce 
qubit-oscillator entanglement which amounts to decoherence.
The relaxation due to the coupling between the oscillator
and the environment will be described by the following master equation 

\begin{figure}
\includegraphics[width=0.45\textwidth]{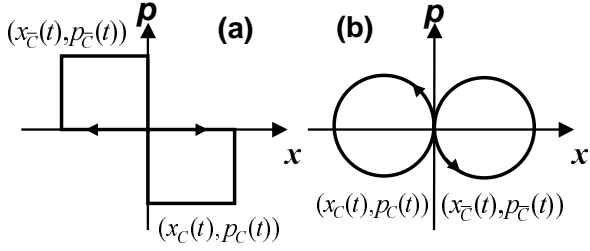}
\caption{The path traversed in phase space of the circuit evolution: (a)
the case of step pulse interaction; (b) the circular case with $\alpha
(t)=\alpha_0\cos \frac{\pi t}{2\tau }+ i\alpha_0\sin \frac{\pi t}{2\tau }$. }
\end{figure}

\begin{eqnarray}
\partial _t\rho &=&-i[H(t),\rho ]+\frac 12\kappa (2a\rho a^{\dagger}-
\{a^{\dagger }a,\rho \})  \nonumber \\ &:= &(\mathcal{L}_0+\mathcal{L}_{th})\rho ,
\label{master} \end{eqnarray}
where $\kappa $ denotes  the relaxation rate. The Liouville operator contains
two terms $\mathcal{L=L}_0+\mathcal{L}_{th}$ with $\mathcal{L}_0\rho
=-i[H(t),\rho ]$. Let us first examine, for transparency, the influence of 
dissipation on the no-jump trajectory, namely, the evolution generated by 
the following non-Hermitian Hamiltonian 
\begin{equation}
\tilde{H}(t)=H(t)-\frac i2\kappa a^{\dagger }a.  \label{hamtra}
\end{equation}
Let us consider the circuit of step pulses applied to the system. If the
relaxation rate $\kappa $ is small, i.e. $\kappa \ll \alpha _0\sim 1/\tau,$
then the non-unitary transformation generated by every pulse, $\tilde{P}
_{\pm }^{x,p}(\tau )$, can be approximately expanded by resorting to the 
Baker-Hausdorf relation 
$e^{A+B}=e^Ae^Be^{-\frac 12[A,B]}e^{-\frac 1{12}([A,[A,B]]+[B,[B,A]])}\cdots.$
In this way it is not difficult to realize that, to the first order 
in $\kappa /\alpha_0$, 
the sequence of four step pulses gives rise to the following transformation 
\begin{eqnarray}
\tilde{U}_C(T) &=&\tilde{P}_{+}^x(\tau )\tilde{P}_{+}^p(\tau )\tilde{P}%
_{-}^x(\tau )\tilde{P}_{-}^p(\tau )  
\label{nonuni} \\
&\approx &e^{-i2\alpha _0^2\tau ^2J_y^2}e^{-\sqrt{2}\kappa \alpha _0\tau
^2xJ_y}e^{\sqrt{2}\kappa \alpha _0\tau ^2pJ_y}e^{-2\kappa \tau a^{\dagger}a}.
\nonumber
\end{eqnarray}
Clearly, the appearance of the middle two factors in the last term of 
Eq. (\ref{nonuni}) leads to the entanglement of qubits with the oscillator 
degrees of freedom, hence results in dependence of the transformation on 
the vibrational state.

The key observation is now to notice that the above fault can be 
effectively suppressed
by performing a further time-reversed pulse sequence.
One can readily obtain that the no-jump trajectory of the time-reversed
pulse sequence will generate the transformation 
\begin{eqnarray}
\tilde{U}_{\bar{C}}(T) &=&\tilde{P}_{+}^x(\tau )\tilde{P}_{+}^p(\tau )\tilde{%
P}_{-}^x(\tau )\tilde{P}_{-}^p(\tau )  
\label{anti} \\
&\approx &e^{-i2\alpha _0^2\tau ^2J_y^2}e^{\sqrt{2}\kappa \alpha _0\tau
^2xJ_y}e^{-\sqrt{2}\kappa \alpha _0\tau ^2pJ_y}e^{-2\kappa \tau a^{\dagger
}a}.
\nonumber
\end{eqnarray}
From Eqs. (\ref{nonuni}) and (\ref{anti}), it can be  seen that the influence of 
dissipation  in  the two circuits  counteract each other. 
Therefore, a gate operation resistant against dissipation can be
enacted by combining these two opposite pulse circuits 
\begin{eqnarray}
\tilde{U}(T_2) &=&\tilde{U}_{\bar{C}}(T_2/2)\tilde{U}_C(T_2/2)  \nonumber \\
&\approx &e^{-i2(\alpha _0\tau )^2J_y^2}e^{-2\sqrt{2}\kappa \tau a^{\dagger
}a},  \label{total}
\end{eqnarray}
where $T_2=\sqrt{2}T$ is the whole time period of the double-loop process.
Since the two factors of the last term of Eq. (\ref{total}) act individually
on the qubits and the oscillator degree of freedom, the dissipation-induced 
dependence of the gate operation on the oscillator state has been removed.

It is important to stress that the cancellation of the unwanted influence of 
dissipation 
achieved by the above procedure is essentially a sort of group-symmetrization 
decoupling process \cite{bang}. 
Notably, in the present system, the enacted ${\bf{Z}}_2$-invariance of the
dynamics under the map $J_y\rightarrow -J_y$ resides in
the geometric feature of the global transformation (\ref{pulse}); 
therefore the devised symmetrized (anti-symmetrized) evolution (\ref{total})
endows the scheme with a two-fold resilience against the noise sources due to 
the coupling
with the oscillator: thermal fluctuations and dissipation.

The analysis carried over up to now was limited to the evolution in
the no-jump trajectory. To get a full picture of the dissipative dynamics, 
i.e. the resolution of the master equation, generally requires to resort to 
a numerical approach. Remarkably, as we shall show in the
following, the master equation of the present system (\ref{master}) can be 
solved analytically by using 
group-theoretic techniques at a super-operatorial level.

Before presenting the solution, it is convenient to introduce the
following notation \cite{wangan}. Let us define the set of left operators $%
O_L^i$ and right operators $O_R^i:$ the left operator action on the density matrix 
is  given, as usual,  by multiplication on left whereas the action of the right operators
is defined as 
\begin{equation}
O_R^1\times O_R^2\times \cdots \times O_R^m\rho =\rho O^m\times \cdots
\times O^2\times O^1.  \label{defin}
\end{equation}
By means  of these notations, the operators contained in the master
equation (\ref{master}) can be conveniently described. For example, the
operator $\mathcal{L}_{th},$ which describes the pure relaxation process, can
be denoted by $\mathcal{L}_{th}=\frac 12\kappa (2K_{-}-K_0+1)$,
where $K_0=a_L^{\dagger }a_L+a_R^{\dagger }a_R$, $K_{-}=a_La_R^{\dagger }$
and $K_{+}=a_L^{\dagger }a_R$. Moreover, in view of the fact that  these (super)operators
generate  a simple $\rm{su}(1,1)$ algebra: $[K_0,K_{\pm }]=\pm 2K_{\pm }$, $%
[K_{-},K_{+}]=K_0$, the process of pure  relaxation is given by 
$e^{\mathcal{L}_{th}t}=e^{-K_{-}}e^{\frac{\kappa t}2(1-K_0)}e^{K_{-}}$. 
The established formalism lead in a rather straightforward fashion to the explicit
description of the relaxation dynamics for a general oscillator coherent state $|\beta
\rangle $ \cite{cen-zan-next}, 
\begin{equation}
e^{\mathcal{L}_{th}t}|\beta \rangle \langle \beta |=|\beta e^{-\frac
12\kappa t}\rangle \langle \beta e^{-\frac 12\kappa t}|.  \label{relax}
\end{equation}

Suppose now that the finite-time evolution  governed by the master equation (\ref{master}) 
is given by a map $\Lambda$, i.e., 
\begin{equation}
\rho (t)=\Lambda (t)\rho (0).  \label{evol}
\end{equation}
Note that a natural picture for the system is provided by  basis of 
$J_y$-eigenstates $|l\rangle $. The time evolution for every element of the
operator $\Lambda (t)$, defined as $\Lambda ^{ll^{\prime }}(t)=\langle
l|\Lambda (t)|l^{\prime }\rangle $, can be obtained: 
\begin{equation}
\partial _t\Lambda ^{ll^{\prime }}(t)=(\mathcal{L}_0^{ll^{\prime }}+\mathcal{%
L}_{th})\Lambda ^{ll^{\prime }}(t),  \label{superope}
\end{equation}
where 
\begin{equation}
\mathcal{L}_0^{ll^{\prime }}=-il(\alpha a_L+\alpha ^{*}a_L^{\dagger
})+il^{\prime }(\alpha a_R+\alpha ^{*}a_R^{\dagger }).  \label{superope0}
\end{equation}
It should be clear that Eq. (\ref{superope}) suggests a linear algebraic
structure, including, besides the aforementioned $\rm{su}(1,1)$ algebra, the
both the left and right Heisenberg algebras  
\{$ a_L,a_L^{\dagger },1$\} and \{$a_R,a_R^{\dagger },1$\}.
According to the  group-theoretic approach developed in Ref. \cite{wangsj}, such a
linear algebraic system is exactly solvable. We summarize below the
solution of Eq. (\ref{superope}), details of the resolution will be reported 
elsewhere \cite{cen-zan-next}. 
\begin{eqnarray}
\Lambda ^{ll^{\prime }}(t) &=&e^{-i\mathcal{A}^{ll^{\prime }}(t)}e^{-il\xi
_{+}^{*}(t)a_L^{\dagger }}e^{-i[l^{\prime }\xi _{+}(t)+\Delta _{ll^{\prime
}}\xi _{-}(t)]a_L}  
\nonumber \\
&&\times e^{il^{\prime }\xi _{+}(t)a_R}
e^{i[l\xi _{+}^{*}(t)-\Delta _{ll^{\prime }}\xi_{-}^{*}(t)]a_R^{\dagger }}
e^{\mathcal{L}_{th}t},
\label{solution}
\end{eqnarray}
where $\Delta _{ll^{\prime }}=l-l^{\prime }$ and the coefficients are given
by 
\begin{eqnarray}
\xi _{\pm }(t) &=&e^{\mp \frac 12\kappa t}\int_0^t\alpha (t^{\prime })e^{\pm
\frac 12\kappa t^{\prime }}dt^{\prime },  \nonumber \\
\mathcal{A}^{ll^{\prime }}(t) &=&i\kappa ll^{\prime }\int_0^t|\xi
_{+}(t^{\prime })|^2dt^{\prime }  \nonumber \\
&&-i\int_0^t[l^2\alpha (t^{\prime })\xi _{+}^{*}(t^{\prime })+l^{\prime
2}\alpha ^{*}(t^{\prime })\xi _{+}(t^{\prime })]dt^{\prime }.~~
\label{coeffi}
\end{eqnarray}
The above analytical solution fully characterizes the decoherence effects induced by
the dissipation of the oscillator mode, e.g. vibrational relaxation in an ion trap. 
The integrals in the functions  $\xi _{\pm }(t)$ might  clearly  be non zero;
therefore, the transformation described by  equation (\ref{solution}) will be inevitably
depend on the oscillator state. More specifically, let us suppose that the oscillator
state  is prepared in a general coherent state $|\beta \rangle $ and the whole system 
is initially prepared in the product state
$\rho (0)=\rho _{in}(0)\otimes |\beta \rangle \langle \beta |$. 
By substituting $\Lambda ^{ll^{\prime }}(t)$ of (\ref{solution}) into equation (\ref{evol})
and tracing over the oscillator degree of freedom, one obtains the
evolution of the reduced density matrix of the  qubits  
\begin{eqnarray}
\rho _{in}^{ll^{\prime }}(t) &=&e^{-i\mathcal{A}^{ll^{\prime
}}(t)}e^{ll^{\prime }|\xi _{+}(t)|^2}  \nonumber \\
&&\times e^{-i\Delta _{ll^{\prime }}(\beta \xi _{-}(t)+\beta ^{*}\xi
_{-}^{*}(t))e^{-\frac 12\kappa t}}\rho _{in}^{ll^{\prime }}(0),
\label{coherent}
\end{eqnarray}
where we have taken the element $\rho _{in}^{ll^{\prime }}(t)$ in the $J_y$
representation.

It should now become evident that the dissipation-induced effects can be suppressed
by the symmetrization procedure. The  application of the time-reversed pulse
circuit $\alpha (t)$ pulse will quench the integrations contained in 
$\xi _{\pm }(t)$. Specifically, note that the polynomial expressions of $\xi
_{\pm }(t)$ is given by  
$\xi _{\pm }(t)=e^{\mp \frac 12\kappa t}\sum_{k=0}^\infty \frac{(\pm )^k}{k!}
I^{(k)}(t),  \label{polyn}
$
where $I^{(k)}(t)=\int_0^t\alpha (t^{\prime })(\frac \kappa 2t^{\prime})^kdt^{\prime }.$
For the step pulse sequence indicated by Eq. (\ref{pulse}),
one has  $I^{(0)}(T)=0$ and nonvanishing terms $I^{(k)}(T)\propto(\kappa T)^k$ 
of $k>0$; but for the time-reversal symmetrized pulse sequence
indicated by Eq. (\ref{total}), there are $I^{(0)}(T_2)=I^{(1)}(T_2)=0$,
leaving only higher order terms of $k>1$. In fact, such a  symmetrization procedure
amounts to a  decoupling, in which the dissipation
effects are removed to the lowest-order in the scale $\kappa /\alpha _0$. 
In principle, it is possible to cancel the decoherence effects to an arbitrary
high order by an iterative application of the above symmetrization
procedure. The $k$-order decoupling pulse sequence to implement the
gate operation  (\ref{pulse}) is illustrated below 
\begin{equation}
U^{(k)}(T_n)=\underbrace{\cdots \cdots U_{\bar{C}}U_CU_CU_{\bar{C}}{
\underbrace{U_CU_{\bar{C}}{\underbrace{U_{\bar{C}}U_C}_{(1)}}}_{(2)}}}_{(k)},
\label{itera}
\end{equation}
where $n=2^k$ is the number of the pulse circuits and $T_n=\sqrt{n}T$ is the
total  time length of the above pulse sequence.

It is worthwhile to point out that besides inducing the mentioned dependence
of the qubit evolution on the oscillator state, dissipation will also result in
a deviation from the desired gate operation. This latter effect is indicated 
in the coefficient $ \mathcal{A}^{ll^{\prime }}(t)$.
Note that this undesired influence of dissipation, being  embedded within the geometric
action, cannot be removed by the above  symmetrization procedure. 
In Fig. 2 we report in detail the fidelities of the gate operation for different 
applications of the pulse sequences as a function of 
the ratio $\kappa /\alpha _0$. Fidelity is defined as 

\begin{figure}
\includegraphics[width=0.45\textwidth]{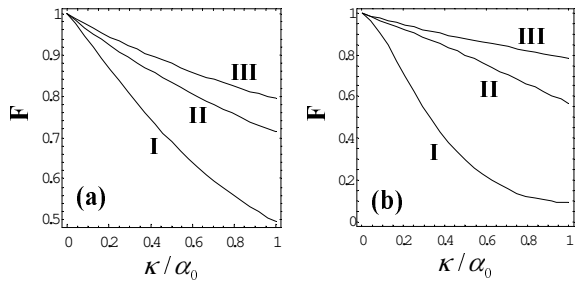}
\caption{The fidelity of the gate operation [Eq. (\ref{fidelity})] for
various applications of step pulse sequences. Curves I, II, and III
correspond respectively to the symmetrization procedure with decoupling
order $k=0,1$ and $2$, in which the number of the pulse circuit is taken 
as $
n=1,2$ and $4$ accordingly. The initial state of the oscillator degree of
freedom assumes (a) $\beta =2$ and (b) $\beta =5$. }
\end{figure}

\begin{equation}
F=\langle \Psi _{\max }|\rho _{in}(T_n)|\Psi _{\max }\rangle ,
\label{fidelity}
\end{equation}
in which $T_n=\sqrt{n\pi }/\alpha _0$ with $n$ the number of the circuits
and $|\Psi _{\max }\rangle =e^{-i\frac \pi 8J_y^2}|00\rangle $ is the fully
entangled state attained by the ideal gate operation acting on the qubit state 
$|00\rangle $.

In summary, we have contrived a novel symmetrization procedure for 
oscillator-assisted geometric quantum gates to suppress the detrimental 
effects induced by dissipation of the
oscillator mode. The existence of the symmetry is related to the invariance 
of this class of geometric operations under the finite-time reversal transformation.
We have devised a time-reversal symmetrization procedure which purports to eliminate
quantum decoherence due to the oscillator dissipative relaxation to an arbitrary high order. 
The resulting quantum computation scheme possesses a two-fold resilient character 
against decoherence: insensitive to the oscillator thermal fluctuations and immune 
to quantum dissipation. Note that a similar symmetrization procedure was involved in
a former refocusing scheme for holonomic quantum computation \cite{cen}.
In comparison, the present work suggests a nonadiabatic manner to implement
a geometrical gate operation robust against both classic and quantum noises. 
Moreover, the resistance against dissipation in the present scheme has been
demonstrated by analytically solving the master equation by means of a powerful  
group-theoretical method at the superoperatorial level.
On the practical side, we expect our results to be especially relevant for
ion-trap quantum computing by geometrical gates.

This work is financially supported by the European Union project TOPQIP
(Contract No. IST-2001-39215).
P.Z. gratefully acknowledges financial support by
Cambridge-MIT Institute Limited.





\end{document}